\documentclass[twocolumn,showpacs,prc,floatfix]{revtex4}
\usepackage{graphicx}
\usepackage{amsmath}
\newcommand{\p}[1]{(\ref{#1})}
\newcommand\beq{\begin{eqnarray}} \newcommand\eeq{\end{eqnarray}}
\newcommand\beqstar{\begin{eqnarray*}} \newcommand\eeqstar{\end{eqnarray*}}
\newcommand{\beqe}{\begin{equation}} \newcommand{\eeqe}{\end{equation}}
\newcommand{\bal}{\begin{align}}

\begin {document}

\title{Superfluid states with moving condensate \protect\\ in
nuclear matter }

\author{ A. A. Isayev}

\affiliation{Kharkov Institute of Physics and Technology,
 Kharkov, 61108, Ukraine}
\date{\today}
\begin{abstract}
Superfluid states of symmetric nuclear matter with finite total
momentum of Cooper pairs (nuclear LOFF phase) are studied with the
use of Fermi--liquid theory in the model with Skyrme effective
forces. It is considered the case of four--fold splitting of the
excitation spectrum due to finite superfluid momentum and coupling
of $T=0$ and $T=1$ pairing channels. It has been shown that at
zero temperature the energy gap in triplet--singlet (TS) pairing
channel (in spin and isospin spaces) for the SkM$^*$ force
demonstrates double--valued behavior as a function of superfluid
momentum. As a consequence, the phase transition at the critical
superfluid momentum from the LOFF phase to the normal state will
be of a first order.  Behavior of the energy gap as a function of
density for TS pairing channel under increase of superfluid
momentum changes from one--valued to universal two--valued. It is
shown that two--gap solutions, describing superposition of states
with singlet--triplet (ST) and TS pairing of nucleons appear as a
result of branching from one--gap ST solution. Comparison of the
free energies shows that the state with TS pairing of nucleons is
thermodynamically most preferable.
\end{abstract}
 \maketitle

  In this report we shall study superfluid states of nuclear
matter with nonzero total momentum of Cooper pairs. First the
states with moving condensate were considered in
Refs.~\cite{LO,FF} with respect to metallic superconductors. In
this case the superconducting condensate has a spatially periodic
structure. Corresponding phase is called the
Larkin--Ovchinnikov--Fulde--Ferrel (LOFF) phase. Recent upsurge of
interest to the LOFF phase is related with the possibility of
formation of this phase in nuclear  \cite{S} and quark matter
\cite{ABR,RSZ}.
  In Ref.~\cite{S}
it was considered in model calculations the case of
neutron--proton superfluidity, when the quasiparticle spectrum is
two--fold split due to asymmetry of nuclear matter and finite
superfluid momentum. It was shown, that the nuclear LOFF phase
appears as a result of a first order phase transition in the
asymmetry parameter from the spatially uniform BCS superconducting
state. Further increase of the asymmetry parameter leads to the
second order phase transition from the LOFF phase to the normal
state. In the study of Ref.~\cite{S} it was assumed that coupling
between isospin singlet and isospin triplet pairing channels can
be neglected.  However, as emphasized in Ref.~\cite{AIP2}, in the
region of low densities coupling between $T=0$ and $T=1$ pairing
channels may be of importance, leading to the emergence of
multi--gap superfluid states, characterized by nonvanishing gaps
in both pairing channels. Thus, we shall consider the case of
four--fold splitting of the quasiparticle spectrum, caused by
account of, first, finite superfluid momentum and, second,
coupling of $T=0$ and $T=1$ pairing channels. Another simplifying
moment in Ref.~\cite{S} is the use of free single particle
spectrum and bare interaction in the gap equation without taking
into account the effects of medium polarization. A
"first--principle" derivation of the pairing interaction from the
bare $NN$ force still encounters many problems, such as, e.g.,
treatment of core polarization~\cite{KR}. Hence, it is quite
natural to develop some kind of a phenomenological theory, where a
phenomenological pairing interaction is employed.  As such a
theory, we shall use the Fermi--liquid (FL) approach~\cite{AKP}.
In the Fermi--liquid model the normal and anomalous FL interaction
amplitudes are taken into account on an equal footing~\cite{AIPY}.
This will allow us to consider consistently influence of the FL
amplitudes on superfluid properties of the nuclear LOFF phase.
Besides, as a potential of NN interaction we choose the density
dependent Skyrme effective forces, used earlier in a number of
contexts for description of superfluid properties both finite
nuclei \cite{DFT,RDN} and infinite  nuclear matter~\cite{SYK,AIP}.

 Superfluid states of nuclear matter are described
  by the normal $f_{\kappa_1\kappa_2}=\mbox{Tr}\,\varrho
  a^+_{\kappa_2}a_{\kappa_1}$ and
 anomalous $g_{\kappa_1\kappa_2}=\mbox{Tr}\,\varrho
a_{\kappa_2}a_{\kappa_1}$ distribution functions of nucleons
($\kappa\equiv({\bf{p}},\sigma,\tau)$,  $\bf p$ is momentum,
$\sigma(\tau)$ is the projection of spin (isospin) on the third
axis, $\varrho$ is the density matrix of the system). We shall
study two--gap superfluid states in symmetric nuclear matter,
corresponding to superposition of states with total spin $S$ and
isospin $T$ of a pair $S=1$, $T=0$ (triplet--singlet (TS) pairing)
and $S=0$, $T=1$ (singlet--triplet (ST) pairing) with the
projections $S_z=T_z=0$ (TS--ST states). In this case, assuming
that a condensate moves with the finite  momentum $\bf q$, the
normal $f$ and anomalous $g$ distribution functions
read~\cite{AIP2,AKP} \bal f_{\kappa_1\kappa_2}&= f_{00}({\bf
p}_1)(\sigma_0\tau_0)_{\kappa_1\kappa_2}\delta_{p_1,p_2},\label{7.2}\\
g_{\kappa_1\kappa_2}&=(g_{30}({\bf p}_1)\sigma_3\sigma_2\tau_2+g_{03}
({\bf
p}_1)\sigma_2\tau_3\tau_2)_{\kappa_1\kappa_2}\delta_{p_1,-p_2+q},
\label{7.3}\end{align}
  where $\sigma_i$ and $\tau_k$ are the Pauli
matrices in spin and isospin spaces, respectively.
 For the energy functional, being invariant with
respect to rotations in spin and isospin spaces, the quasiparticle
energy and the order parameter have the similar structure \bal
\varepsilon_{\kappa_1\kappa_2}&= \varepsilon_{00}({\bf
p}_1)(\sigma_0\tau_0)_{\kappa_1\kappa_2}\delta_{p_1,p_2},\label{8.2}\\
\Delta_{\kappa_1\kappa_2}&=(\Delta_{30}({\bf
p}_1)\sigma_3\sigma_2\tau_2+\Delta_{03}({\bf
p}_1)\sigma_2\tau_3\tau_2)_{\kappa_1\kappa_2}\delta_{p_1,-p_2+q}
\label{8.3}\end{align}
 If to take into account the antisymmetry properties
 $\Delta^{\text{T}}=-\Delta$, $g^{\text{T}}=-g$
 and to set
${\bf p}_1={\bf p}+{\bf q}/2$, ${\bf p}_2=-{\bf p}+{\bf q}/2$
($\bf q$ is total momentum of a pair, $\bf p$ is momentum of one
of nucleons in the center of mass frame of a pair), then  we
obtain\beq \Delta_{30}({\bf p}+\frac{\bf
q}{2})&=&\Delta_{30}(-{\bf p}+\frac{\bf q}{2})\equiv
\Delta_{30}({\bf p},{\bf q}),\\ \Delta_{03}({\bf p}+\frac{\bf
q}{2})&=&\Delta_{03}(-{\bf p}+\frac{\bf q}{2})\equiv
\Delta_{03}({\bf p},{\bf q})\nonumber\eeq and analogous
relationships hold for the functions $g_{30},g_{03}$. Further we
shall write the self--consistent equations for the quantities $
\Delta_{30}({\bf p},{\bf q})$, $\Delta_{03}({\bf p},{\bf q})$.
Using the minimum principle of the thermodynamic potential and
procedure of block diagonalization \cite{AKP}, one
 can express evidently the distribution functions $f_{00},g_{30},g_{03}$
 in
terms of the quantities $\varepsilon$ and $\Delta$: \bal
f_{00}&=\frac{1}{4}\left[(1+n_+^+-n_+^-)
-\frac{\xi_s}{E_+}(1-n_+^+-n_+^-)\right. \label{13} \\
&\quad+\left.(1+n_-^+-n_-^-)
-\frac{\xi_s}{E_-}(1-n_-^+-n_-^-)\right],\nonumber\\
g_{30}&=-\frac{\Delta_+}{4E_+}(1-n_+^+-n_+^-)-
\frac{\Delta_-}{4E_-}(1-n_-^+-n_-^-),\label{11}\\
g_{03}&=-\frac{\Delta_+}{4E_+}(1-n_+^+-n_+^-)+
\frac{\Delta_-}{4E_-}(1-n_-^+-n_-^-).\label{12} \end{align} Here
$f_{00}=f_{00}({\bf p}+ \frac{\bf q}{2})$, $n^\pm=n(\pm{\bf
p}+\frac{\bf q}{2})$ and\begin{gather*}
E_\pm=\sqrt{\xi_s^2+|\Delta_\pm|^2},\quad\Delta_\pm=\Delta_{30}\pm\Delta_{03},\\
\xi_s({\bf p}+\frac{\bf q}{2})=\frac{1}{2}\left(\xi({\bf
p}+\frac{\bf
q}{2})+ \xi(-{\bf p}+\frac{\bf q}{2})\right),\;\\
\xi_a({\bf p}+\frac{\bf q}{2})=\frac{1}{2}\left(\xi({\bf
p}+\frac{\bf q}{2})- \xi(-{\bf p}+\frac{\bf q}{2})\right),\\
\xi({\bf p})=\varepsilon_{00}({\bf p})-\mu^0,\;n_{\pm}=\{\exp
Y_0(\xi_a+E_\pm)+1\}^{-1},\end{gather*}   $\mu^0$ is chemical
potential, which should be determined from the normalization
condition \beq \frac{4}{\cal V}\sum_{\bf p}f_{00}({\bf
p}+\frac{\bf q}{2})=\varrho,\label{13.1}\eeq  $\varrho$  is density of
symmetric nuclear matter. As follows from the structure of the
distribution functions $f_{00}, g_{30}, g_{03}$, the quantity
$$\omega_{\pm,\pm}=\xi_a(\pm {\bf p}+\frac{\bf q}{2})+E_\pm,$$
being the exponent in Fermi distribution functions $n_\pm(\pm {\bf
 p}+\frac{\bf q}{2})$, plays the role of the quasiparticle excitation
 spectrum.
 In the considering case the spectrum is four--fold split due to 1) finite
 superfluid momentum (${\bf q}\not=0$), 2) coupling of TS and ST pairing
 channels ($\Delta_{30}\not=0,\Delta_{03}\not=0$).

To obtain the closed system of equations for the quantities $\Delta$
and $\xi$, it is necessary to set the energy functional of the
system.  In the case of symmetric nuclear matter with TS and ST
pairings of nucleons the energy functional   is characterized by
one normal $U_0$  and two anomalous $V_1,V_2$ FL amplitudes
\cite{AIP2,AIP}. Then one can obtain the self--consistent equations
in the form \bal \xi({\bf p})&=\varepsilon_0^0({\bf
p})-\mu^0+\tilde\varepsilon_{00}({\bf p}),\;\;
\varepsilon_0^0({\bf p})=\frac{{\bf
p}^{\,2}}{2m_{0}},
\label{15}\\
\tilde\varepsilon_{00}({\bf p})&=\frac{1}{2\cal V}\sum_{\bf p'}U_0({\bf
k})f_{00}({\bf p'}),\; {\bf k}=\frac{{\bf  p}-{\bf
p'}}{2},\nonumber\\
 \Delta_{30}({\bf p},{\bf q})&=\frac{1}{\cal V}\sum_{\bf p'}V_1({\bf p},
 {\bf
p}')g_{30}({\bf p}',{\bf q}),\;\label{16}\\
\Delta_{03}(\bf p,{\bf q})&=\frac{1}{\cal V}\sum_{\bf p'}V_2({\bf
p},{\bf p'})g_{03}({\bf p}',{\bf q}),\label{17}\end{align} where
$m_{0}$ being the mass of a bare nucleon.

 Further for obtaining
numerical results we shall use the Skyrme effective interaction.
In the case of Skyrme forces the normal and anomalous FL
 amplitudes read~\cite{AIP} \bal
U_0({\bf k})&=6t_0+6t_3\varrho^\beta\label{17.1}\\
&\quad+\frac{1}{\hbar^2}[6t_1+2t_2(5+4x_2)]{\bf k}^{2}\equiv
d_0+e_0{\bf
k}^{2}\nonumber\\ V_{1,2}({\bf p},{\bf p}')&=t_0(1\pm x_0)
+\frac{1}{6}t_3\varrho^\beta(1\pm x_3)\label{17.2}\\
&\quad+\frac{1}{2\hbar^2}t_1(1\pm x_1)({\bf p}^{2}+{\bf p}'{}^{2}),
\nonumber\end{align} where $t_i,x_i,\beta$ are some phenomenological
constants, characterizing the given parametrization of Skyrme
forces
 (we shall use the  SkP
\cite{DFT}, SkM$^*$ \cite{BGH} potentials). With account of
Eq.~\p{17.1} we obtain \bal &\xi_s({\bf p}+\frac{\bf
q}{2})=\frac{p^2}{2m_1}+\frac{q^2}{8m_2}-\mu, \\
\intertext{where}
\frac{\hbar^2}{2m_{1,2}}&=\frac{\hbar^2}{2m_{0}}\pm\frac{\varrho}{16}
[3t_1+t_2(5+4x_2)]\label{18}\end{align} and the effective chemical
potential $\mu$ should be determined from the normalization
condition \p{13.1}. Besides, expression for the quantity $\xi_a$
reads \beq \xi_a({\bf p}+\frac{\bf q}{2})=\frac{{\bf
pq}}{2m_0}-\frac{e_0}{4} \sum_{{\bf p}'}f_{00}({\bf p}'+\frac{{\bf
q}}{2}){\bf pp}'\label{19}\eeq The normal distribution function
$f_{00}$ in turn depends on the quantity $\xi_a$ and, hence,
expression \p{19} represents an equation for determining the
quantity $\xi_a$. Since the second term in Eq.~\p{19} is
proportional to the scalar product $\bf pq$, solution of
Eq.~\p{19} should be found in the form $\xi_a({\bf p}+\frac{\bf
q}{2})={\bf pq}/2m^*$, where $m^*$ is some effective mass. Using
Eqs.~\p{13}--\p{12}, we present equations for the order parameters
$\Delta_{30},\Delta_{03}$, effective chemical potential $\mu$ and
effective mass $m^*$ as
 \bal \Delta _{30}&=-\frac {1}{4\cal
V}\sum\limits_{{\bf p}' }^{}
 V_1({\bf
p},{\bf p}' ) \left\{ \frac{\Delta_+'}{E_+'}
\biggl(\tanh\frac{\omega_{++}'}{2T}\right.
 \label{21}\\  &\quad+\left.
\tanh\frac{\omega_{-+}'}{2T}\biggr) +
\frac{\Delta_-'}{E_-'}\biggl(\tanh\frac{\omega_{+-}'}{2T}+
\tanh\frac{\omega_{--}'}{2T}\biggr) \right\} , \nonumber\\
 \Delta _{03}&=-\frac {1}{4\cal V}\sum\limits_{{\bf p}' }^{}
 V_2({\bf
p},{\bf p}' ) \left\{ \frac{\Delta_+'}{E_+'}
\biggl(\tanh\frac{\omega_{++}'}{2T}\right.
 \label{22}\\  &\quad+\left.
\tanh\frac{\omega_{-+}'}{2T}\biggr) -
\frac{\Delta_-'}{E_-'}\biggl(\tanh\frac{\omega_{+-}'}{2T}+
\tanh\frac{\omega_{--}'}{2T}\biggr)
\right\} , \nonumber\\
 &\frac{1}{\cal V}\sum_{\bf p}\left\{2-\frac{\xi_s}{2E_+}\biggl(
\tanh
\frac{ \omega_{++}}{2T}+\tanh\frac{ \omega_{-+}}{2T}\biggr)\right.
\nonumber\\
&\quad\left.-\frac{\xi_s}{2E_-} \biggl( \tanh \frac{
\omega_{+-}}{2T}+\tanh\frac{ \omega_{--}}{2T}\biggr)
\right\}=\varrho,\label{23}\\
&\frac{{\bf pq}}{m_0}+\frac{e_0}{16} \sum_{{\bf p}'}{\bf pp}'
\left\{ \biggl( \tanh
\frac{ \omega_{++}'}{2T}-\tanh\frac{ \omega_{-+}'}{2T}\biggr)\right.
\nonumber\\
&\quad\;\;\left.+\biggl( \tanh \frac{ \omega_{+-}'}{2T}-\tanh\frac{
\omega_{--}'}{2T}\biggr) \right\} = \frac{{\bf pq}}{m^*}
                             \label{24}
\end{align} Here
$$
\Delta_\pm'=\Delta_\pm({\bf p}',{\bf q}),\quad
\omega_{\pm,\pm}'=\pm\frac{{\bf p}'{\bf q}}{2m^*}+E_\pm({\bf
p}',{\bf q})$$

Eqs.~\p{21}--\p{24} describe two--gap superfluid states of
symmetric nuclear matter with moving condensate and contain
one--gap solutions with $\Delta_{30}\not=0,\Delta_{03}\equiv0$ (TS
pairing) and $\Delta_{30}\equiv0,\Delta_{03}\not=0$ (ST pairing) as
some particular cases. We shall analyze Eqs.~\p{21}--\p{24} using
the simplifying assumption, that  FL amplitudes $V_1,V_2$ are not
equal to zero only in a narrow layer near the Fermi surface:
$|\xi_s|\le\theta,\, \theta\ll\varepsilon_F$ (further we set
$\theta=0.1\varepsilon_F$).

\begin{figure}[htbp]
\includegraphics[height=14.0cm,width=8.6cm,trim=49mm 66mm 53mm 51mm,
draft=false,clip]{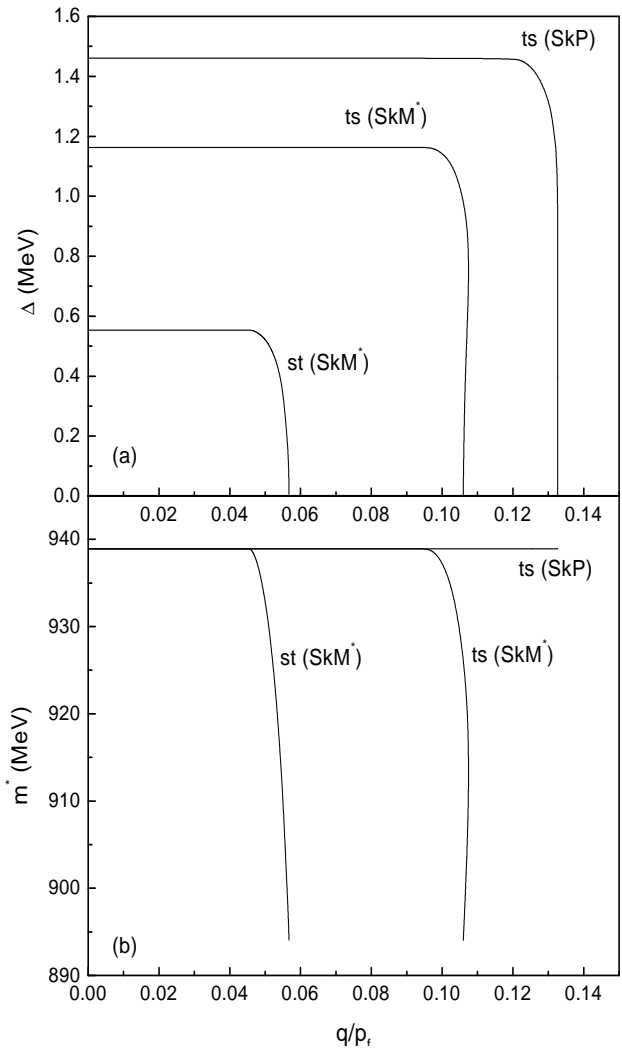}
\caption{Energy gap  (top) and  effective mass $m^*$ (bottom) vs.
superfluid momentum for different types of pairing and Skyrme
forces at $\varrho=0.03\ \mbox{fm}^{-3}$ and zero temperature.
}\label{fig1}\end{figure}

First  we shall find the dependence of the order parameters
$\Delta_{30}(p=p_{\text{F}}),\Delta_{03}(p=p_{\text{F}})$ from the
superfluid momentum at zero temperature. We begin our analysis
with finding one--gap solutions of the self--consistent equations.
Results of numerical determination of the energy gap are presented
in the Fig.~\ref{fig1}a.
 It is seen, that a general tendency is
quite clear: at low superfluid momenta the energy gap retains its
constant value and then rapidly decreases and vanishes at some
critical point. This means, that one--gap superfluid states with
moving condensate will disappear at large enough superfluid
momenta. However, for TS pairing of nucleons, interacting via
SkM$^*$ effective potential, the phase curve has an interesting
peculiarity, namely,  the energy gap demonstrates nontrivial
double--valued behavior in the region, where it sharply falls.
Such behavior differs from the ordinary one--valued behavior of
the energy gap, as, e.g., in the case of TS pairing and SkP
effective interaction.  To understand this difference, we have
determined the effective mass $m^*$ as a function of superfluid
momentum,~Fig.~\ref{fig1}b. For the SkP potential the mass $m^*$
is equal to the bare mass $m_0$ of a nucleon practically for all
superfluid momenta. Unlike to this behavior, the effective mass
$m^*$ for the SkM$^*$ potential rapidly decreases close to the
region of phase transition to the normal state. Since the mass
$m^*$ enters into Eqs.~\p{21}--\p{24} only through the ratio
$q/m^*$, descent of the effective mass leads to the increase of
the effective shift between the centers of proton and neutron
Fermi spheres, on which the paired proton and neutron lie. This
gives the possibility to the appearance of the second solution for
the energy gap. According to Eq.~\p{17.1}, the parameter $e_0$ of
the normal FL amplitude $U_0$ determines the sign before  the sum
in the l.h.s. of Eq. \p{24}, and, hence, determines whether the
mass $m^*$ will be greater (if $e_0<0$) or less (if $e_0>0$), than
the bare mass $m_0$ (if corresponding  sum  is nonzero). For the
SkM$^*$ potential $e_0>0$ while for the SkP potential $e_0<0$,
that explains the difference in the behavior of the effective mass
for these two forces.  However, diminishing behavior of the
effective mass $m^*$ does not guarantee that the energy gap will
have two--valued behavior as a function of superfluid momentum. In
Fig.~1a we have plotted also the dependence of the energy gap  for
ST pairing of nucleons and SkM$^*$ potential. As follows from
here, in spite of correctness of the inequality $e_0>0$, the
energy gap demonstrates the usual one--valued behavior.  Since ST
coupling constant in the Skyrme model is always less then TS one
\cite{AIP2},  one can conclude, that the pairing interaction should
be  strong enough for the second solution to be developed. The
possible two--valued behavior of the energy gap has important
consequence. If to go from the region of large enough superfluid
momenta in the direction of low momenta, then at some critical
point the energy gap will arise by a jump, and, hence, the phase
transition to the superfluid phase will be of a first  order.

\begin{figure}[htbp]
\includegraphics[height=7.0cm,width=8.6cm,trim=51mm 120mm 51mm 69mm,
draft=false,clip]{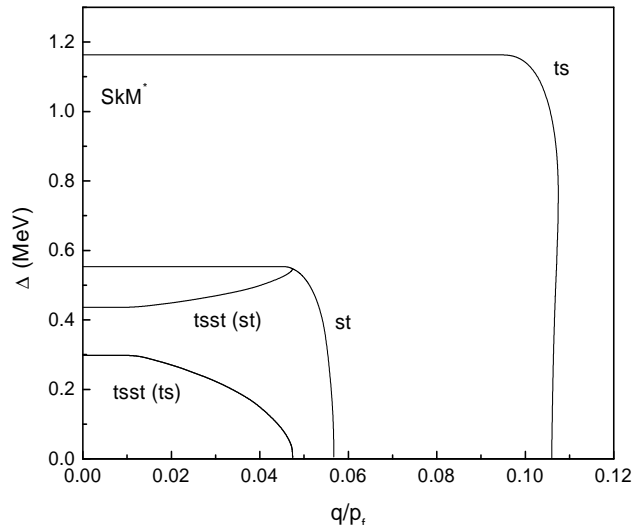} \caption{Order parameters
$\Delta_{30},\Delta_{03}$ vs. superfluid momentum for SkM$^*$
force at $\varrho=0.03\ \mbox{fm}^{-3}$ and zero temperature.
}\label{fig3}\end{figure}

Now we go to the study of two--gap superfluid states when both
order parameters, $\Delta_{30}$ and $\Delta_{03}$, are not equal to
zero.    Results of numerical determination of the order
parameters $\Delta_{30}(q),\Delta_{03}(q)$ on the base of Eqs.
\p{21}--\p{24}  are presented in Fig.~\ref{fig3}.
   Here $tsst(ts)$ and $tsst(st)$ are notations for the
dependencies of TS and ST order parameters in the TS--ST solution
of the self--consistent equations.  As seen from Fig.~\ref{fig3},
TS--ST solutions appear as a result of branching from  one--gap ST
solution  (in the branching point
$\Delta_{30}=0,\Delta_{03}=\Delta_{03}^{st}$, $\Delta_{03}^{st}$
being one--gap ST solution).  Note that the self--consistent
equations have two--gap solutions in the case of SkM$^*$
potential, but have no such solutions for the SkP potential. As
clarified in the Ref.~\cite{AIP2}, for the existence of TS--ST
solutions it is necessary that  TS and ST
 coupling constants  must be of the same
order of magnitude. However, this condition is broken for the
SkP potential, where TS coupling constant is much larger than
ST one.

\begin{figure}[htbp]
\includegraphics[height=7.0cm,width=8.6cm,trim=51mm 120mm 51mm 69mm,
draft=false,clip]{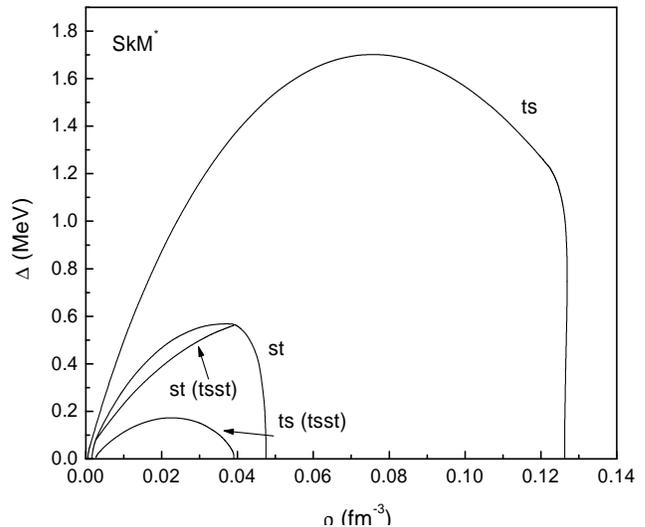} \caption{Order parameters
$\Delta_{30},\Delta_{03}$ vs. density for SkM$^*$ force at
$q/p_F=0.04$ and zero temperature. }\label{fig6}\end{figure}

Since we use density dependent effective interaction, it allows us
to study
 also the dependence of the order parameters from density of nuclear
 matter. The results of numerical determination of one--gap and
two--gap solutions of self--consistent equations at fixed
superfluid momentum  are presented in Fig.~\ref{fig6}.
 As
one can see, superfluidity with finite superfluid  momentum exists
in finite density region $(\varrho_1(q),\varrho_2(q))$, excluding
some vicinity of the point $\varrho=0$ (for TS pairing the left
point
 $\varrho_1(q)$ is very close to  $\varrho=0$).   The most
 important peculiarity, i.e., the double--valued behavior of the energy gap
 for TS pairing in the case of SkM$^*$ potential is preserved for
 the given ratio $q/p_{\text{F}}$.
 For other types of
 pairing and Skyrme forces, considered earlier (including the SkP
potential, which is not shown in Fig.~\ref{fig6}), the order
parameter for one--gap solutions has one--valued behavior.  In the
case of two--gap solutions
 the mechanism of their appearance is
  similar to the considered above, i.e., it is branching from
   one--gap ST solution.

\begin{figure}[htbp]
\includegraphics[height=7.0cm,width=8.6cm,trim=51mm 120mm 51mm 69mm,
draft=false,clip]{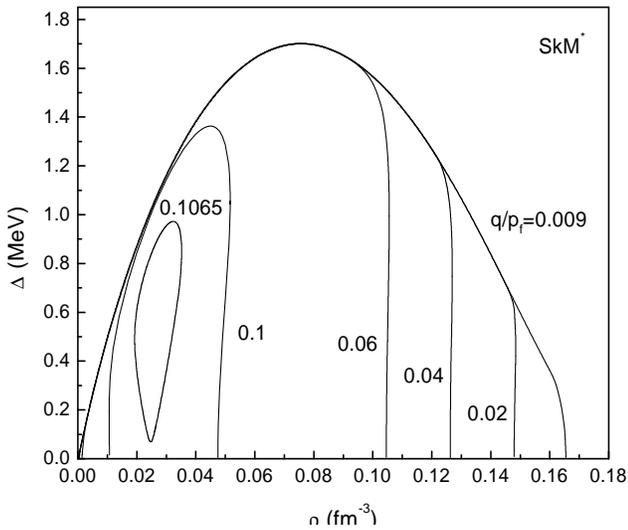} \caption{Energy gap vs. density in the
case of TS pairing and SkM$^*$ force for different values of the
ratio $q/p_F$. }\label{fig5}\end{figure}

However, in general case behavior of the energy gap as a function
of density  is more complicated. In Fig.~\ref{fig5} we plot the
dependence of the energy gap in the case of TS pairing and SkM$^*$
interaction for the set of fixed values of the ratio $q/p_F$.
 It is seen, that behavior
of the phase curves may be
 one--valued or two--valued, that depends on the value of the ratio $q/p_F$.
At small enough superfluid momentum (e.g., for $q=0.009p_f$) the
gap behaves as a one--valued function of density; for the ratios
$q/p_F$, larger than some critical value $q_1/p_F$, the gap has
two--valued behavior in the region close to the right critical
 point $\varrho_2(q)$. Further increase of the ratio $q/p_F$ leads to
 formation of the second part of the phase curve with double--valued
 behavior in the region close to the left critical point $\varrho_1(q)$
 (e.g., for $q=0.1p_f$).  When $q$ increases, the
 regions with double--valued behavior of the gap begin to approach and
 at $q=q_c$ ($q_c\approx0.106 p_F$) it takes place contiguity of the
 regions with two solutions. For  $q>q_c$ the phase curves are
 separated from the density axis and turn into the closed oval curves.
 Under further increase of $q$ the dimension of the oval curves is
 reduced and at some $q=q_m$ the oval curves shrink to a point
 ($q_m\approx0.108p_F$).  For the values $q>q_m$ TS superfluidity of
 nuclear matter vanishes.  Thus, in the range $0<q<q_1$ the gap is
 one--valued function of density, in the range $q_1<q<q_2$ the phase
 curve $\Delta_{30}=\Delta_{30}(\varrho)$ has one part with double--valued
 behavior,  for $q_2<q<q_{c}$ it contains two distinct parts with
 double--valued behavior and for $q_c<q<q_{m}$ the gap has a universal
 double--valued behavior.

 Since we have a few solutions of self--consistent equations, it
 is necessary to check, which solution is thermodynamically
 favorable. Calculations show that due to the large size of the
 gap in TS pairing channel the free energy of the
 corresponding state  as a function of momentum or density much
 smaller then for the case of ST and TS--ST pairing. Hence, TS
 superfluid state wins competition for the thermodynamic
 stability. If to compare the free
energies of two different branches, corresponding to
double--valued behavior of the energy gap in TS pairing channel,
then the branch with larger size of the gap will be
thermodynamically
 favorable.

In summary, we studied superfluidity of symmetric nuclear matter
with moving condensate in the FL model with density dependent
Skyrme effective interaction (SkM$^*$, SkP forces). It has been
considered the case, when the quasiparticle excitation spectrum is
four--fold split due to finite superfluid momentum and coupling of
$T=0$ and $T=1$ pairing channels. Apart from the renormalization
of the chemical potential and bare mass of a nucleon, taking into
account of the normal FL amplitude leads to appearance of
additional effective mass $m^*$ in the linear on superfluid
momentum term in single particle energy.
 It is shown
that at zero temperature the energy gap in TS pairing channel for
SkM$^*$ potential demonstrates two--valued behavior as a function
of superfluid momentum. This is caused by the decreasing behavior
of the effective mass $m^*$ close to the region  of the phase
transition to the normal state and strong enough interaction in TS
pairing channel. The behavior of the energy gap as a function of
density in TS pairing channel in general case is more complicated
and under increase of  superfluid momentum it changes from
one--valued to universal two--valued character (until
superfluidity disappears at some critical momentum).  Two--gap
solutions of self--consistent equations, corresponding to the case
when both TS and ST order parameters  are not equal to zero,
appear as a result of branching from one--gap ST solution.

Calculation of the free energy shows that TS superfluid state is
thermodynamically most preferable state.  In the case of
double--valued behavior the gap changes in the critical point by a
jump and, hence, the phase transition from the LOFF phase to the
normal state will be of a first order (in superfluid momentum or
density). Since the possible two--valued behavior of the gap will
be preserved for small asymmetry, this will be also true for
weakly asymmetric nuclear matter, that differs from the picture of
a second order phase transition, considered in \cite{S}.

The author is grateful for discussions and useful comments to S.
Peletminsky, G. Roepke, H.-J. Schulze and A. Yatsenko. The
financial support of STCU (grant No. 1480) is acknowledged.

   \end{document}